# SINGLE SUPPLY OF ELECTRICAL ENERGY IN THE SUPPLY DISTRIBUTION MODEL: USING GAMS TO MODEL THE EFFECTS OF NETWORK PARAMETERS ON THE LEVEL OF CONSUMER DEMAND

Hamid R.Safarpour[1] (M.S. Student), Amirhossein Fathi[2] (Ph.D. Student)
Kevin M. Hubbard[1], Ph.D. and Emmanuel S. Eneyo[1], Ph.D.

[1]Department of Mechanical and Industrial Engineering

Southern Illinois University of Edwardsville

Edwardsville, Illinois, USA

hsafarp@siue.edu, khubbar@siue.edu, eeneyo@siue.edu

[2]Department of Energy Systems
Sharif University of Technology
Tehran, Iran
Amirhossein_fathi@ieee.org

**Abstract**

*The proposed model of this study is a single supply of electrical energy and is used in distribution systems. The objective of this study is to optimize the distribution of active and reactive energy in the supply of sub-distribution or the market. The model proposed in this paper accounts for the effects associated with switching capacitors on distribution and dispatching centers. Using capacitors in the power grid has an effect on bus voltage and consequently the amount of active and reactive energy demand of the consumer. The optimized supply is determined through the solution of a mathematical operation model. In this study, the effect of voltage changes on consumer demand for residential, commercial and industrial power are quantified and embodied in an analytical model, which is then solved using General Algebraic Modeling System (GAMS) software.*

# I. Introduction

Modeling power consumption under the assumption of constant power in terms of network variables and disregarding the effect of network parameters has low accuracy; the low voltage change on the consumer's bus has negligible effect on the amount of energy consumption. This incorrect modeling leads to inaccuracies in power supply analysis and prediction. The use of a more comprehensive model with the capability of taking into consideration the bus voltage is needed. Conventional models such as constant active power, constant impedance, and the constant current and exponential models are some types of consumers which the effect of voltage on their consumption energy levels has been observed. New approaches have been offered for a more accurate modeling of consumption. Variable definition of the model is shown in Table 1.

The models that network variables such as voltage and frequency have been found in them. The complexity of the new models, conventional methods for solving them will not convex into an answer; since the results of these models are based on trial and error so the ability to make the model of all consumers is not possible. Therefore, this consumer model has been expressed with regard to this problem. In this model, voltage quality over consumption energy of loads has been considered. Since distribution companies in this final model are considered private corporations, so in this regard the optimum utilization of supply purpose is to maximize profits. Distribution of the proposed model will determine the demand of the market for the next day market, the voltage quality, and capacitor switching will be determined. The next day market is a market for supply planning for any time of the future day that will be formed with regard to load flow constraints. Unlike its name, it is usually a forecast which is done a few days before the actual date. Supply and demand in the market for each hour of t the next day have been presented in their proposals. In this model it is assumed the cost per unit of active and reactive power is independent of the demand model.

TABLE 1: VARIABLES DEFINITION

| Variable | Description | Type of Variable | Variable | Description | Type of Variable |
|---|---|---|---|---|---|
| $TotalRevenue$ | Total Revenue of distribution company in 24 hrs | Endogenous | $q(t, type, i)$ | Predicted Reactive Energy per $MVArh$ for the final consumer at bus i, group type ,and Time t | Endogenous |
| $TotalCost$ | Total Cost of distribution company in 24 hrs | Endogenous | $p_{pu}(t, type, i)$ | predicted active energy Per $MWh$ for the final consumer at bus i , group type ,and time t | Endogenous |
| $Profit$ | Total Profit distribution company in 24 hrs | Endogenous | $q_{pu}(t, type, i)$ | Predicted Reactive Energy per $MVArh$ for the final consumer at bus i, group, type and time t | Endogenous |
| $RC(t, i)$ | Income from the sale of electricity to consumers in bus i at time t | Endogenous | $PM(t)$ | Purchase cost per unit of reactive power from the wholesale market in terms of $\frac{\$}{MWh}$ at time t | Exogenous |
| $ca(t, type)$ | Active energy per unit ($\frac{\$}{MWh}$)sales price to the final consumer in group type and time t | Exogenous | $QM(t)$ | Purchase cost per unit of reactive power from the wholesale market in $\frac{\$}{MVArh}$ terms of at time t | Exogenous |

| Variable | Description | Type of Variable | Variable | Description | Type of Variable |
|---|---|---|---|---|---|
| $cr(t, type)$ | Reactive energy per unit sales price to the final ($\frac{\$}{MVArh}$) consumer in type and time t | Exogenous | $CM(t)$ | Purchase cost from the wholesale market at time t | Endogenous |
| $p(t, type, i)$ | Predicted Active Energy Per MWh for the final consumer at bus i, group, type and time t | Endogenous | $V_{pu}(i)_{min}$ | The minimum permissible bus voltage in bus i in Per unit | Exogenous |
| **Variable** | **Description** | **Type of Variable** | **Variable** | **Description** | **Type of Variable** |
| $V_{pu}(i)_{max}$ | The maximum permissible bus voltage in bus i in Per unit | Exogenous | $Q_{Slack}^{Neg}(t)$ | the proposed reactive power for sale to an external networks or a market at time t | Endogenous |
| $V_{pu}(i)_{max}$ | The maximum permissible bus voltage in bus i in Per unit | Exogenous | $Q_{Slack}^{Pos}(t)$ | the proposed reactive power for purchase from an external networks or a market at time t | Endogenous |
| $V_{pu,Slack,Constant}$ | Voltage of Slack per unit | Exogenous | $P_{Slack}(t)$ | Traded Active Energy in distribution companies to wholesale market at time t | Endogenous |
| $V_{pu}(t, i)$ | Voltage of Bus i at time t in per unit | Endogenous | $Q_{Slack}(t)$ | Traded Reactive Energy in distribution companies to wholesale market at time t | Endogenous |
| $Y_{pu}(i, j)$ | Size of Entries in row i and column j of admittance matrix based on Per unit | Endogenous | $P_{Slack}^{Max}(t)$ | Maximum reactive power can be purchased from the wholesale market at time t | Exogenous |
| $\delta(t, i)$ | Voltage angle of bus i at time t | Exogenous | $Q_{Slack}^{Max}(t)$ | Maximum reactive power can be purchased from the wholesale market at time t | Exogenous |
| $\theta(i, j)$ | Angle element in row i and column j of the matrix Admittance | Exogenous | $NCap(i, t)$ | Number of capacitors connected to bus i at time t | Endogenous |
| $f$ | System frequency | Exogenous | $NCap_{max}(i)$ | Maximum number of capacitor on bus i | Exogenous |
| $S_{pu,cap}(i, t)$ | Injected reactive power of bus i at time t in per unit | Endogenous | $P_{Slack}^{Neg}(t)$ | Value of Proposed Active Energy for sale from an external grid or a market at time t | Endogenous |
| $C_{pu,Base}(i)$ | Capacitors connected to bus i at time t based on per unit | Exogenous | $P_{Slack}^{Pos}(t)$ | Value of Proposed Active Energy for purchase to an external grid or a market at time t | Endogenous |

## II. The model

The proposed model is an energy supply model, and therefore consists of the objective function and constraints. In Section II-A, the objective function and the method which the objective model was defined are discussed and in section II-B, the constraints and the method which the model was modeled are discussed.

### A. Objective Function

With the renewal of the power structure in the main stimulus to motivate companies to reduce losses of distribution network, earn higher profits because the companies of distribution has a private nature, so the company will aim to increase profits and decrease losses of the network therefore, increase the final cost of electricity distribution companies to increase compelling interest to optimize utilization of network, and therefore it results in reducing losses. However, according to the aim of companies, the minimum loses does not guaranteed to have maximum profit. Therefore, the proposed model objective function is to maximize total profits in 24 hours forecasting window (Garg and Sharma 2008, Kirschen et. al 2004). Distribution of profit per hour was a total difference in cost and revenue, and it is obtained using equation (1) model.

$$Profit = Total\ Revenue - Total\ Cost \qquad (1)$$

The source of Income for distribution companies in an hour, which is selling electricity to consumers using equation (2), is calculated.

$$Total\ Revenue = \sum_{t=1}^{24}\sum_{i=2}^{N} RC(t,i) \qquad (2)$$

Income from energy sales to consumers is comprised of two parts. The first is the energy sale of active power and the second is reactive power sale. Energy cost per hour for each group of consumers can be different. In this model it is assumed that each bus has a same type of consumption. Using equation (3) the cost for the bus "i" and time" t" is modeled.

$$RC(t,i) = ca(t,type) * p(t,type,i) + cr(t,type) * q(t,type,i) \qquad (3)$$

Distribution cost in the electricity market is a time cost of purchasing, and using equation (4) is modeled.

$$TotalCost = \sum_{t=1}^{24} CM(t) \qquad (4)$$

In this model, it is assumed that the purchase cost per unit of electrical energy market, is an independent variable. Purchase cost of the network at any time using equation (5) is modeled. This relationship consists of two cost functions, the first part is associated with purchase cost of the active power, and the second part is associated with purchase cost of the reactive power.

$$CM(t) = P_{Slack}^{Pos}(t) * PM(t) + Q_{Slack}^{Pos}(t) * QM(t) \qquad (5)$$

*B. Constraints*

The constraints used here include load flow constraints, each bus voltage, and performance capacitor in network. Usually the node voltage method that is most suitable for power system analysis is used for formulation of network equations. This type of load flow is used for calculating the exact amount of the loss voltage in the distribution lines and also for increasing the accuracy of calculation.

For modeling constraints equation (6) and (7) are used. These equations represent the balance between

$$-\frac{q_{pu}(t,type,i)}{1\ hr} + \frac{Q_{pu,cap}(i,t)}{1\ hr} = \sum_{j=1}^{NB} V_{pu}(t,i) * V_{pu}(t,j) * \sin(\delta(t,i) - \delta(t,j) - \theta(i,j)) \quad (6)$$

$$-\frac{p_{pu}(t,type,i)}{1\ hr} = \sum_{j=1}^{N} V_{pu}(t,i) * V_{pu}(t,j) * Y_{pu}(i,j) * \cos(\delta(t,i) - \delta(t,j) - \theta(i,j)) \quad (7)$$

generation and consumption.

In Equation 6 & 7, Y represents an impedance of the line. Distribution Lines based on the voltage and the type (aerial, cable, etc.) will be classified in three models that include:

- short distribution lines
- medium distribution lines
- long distribution lines

If the line length is less than about 80 km (50 miles) or the voltage is not more than 69 KV, the line capacitance can be neglected without much error, and the short line model can be used. Since the length of the distribution lines is less than 80 km and the distribution voltage level is less than kV 69, so short-line distribution model is used to in this study.

a) *Classification of Load Consumption:*

In this model, three categories of consumers, industrial, commercial and residential have been divided.

b) *Distribution Load Modeling:*

Typically frequently changing voltage causes the power consumption of most of loads to be changed. Hence the constant power load model can reduce the accuracy of the model (Liu et.al 2002 and Singh et.al 2009).

In this paper to model the relationship between active power and reactive power of load with its voltage, equations (8) and (9) are used. Also it is possible with these equations to classify the loads into one of these four groups: constant current (I), constant impedance (Z), constant power (P&Q) and the exponential model (Saadat 2002).

Constant Current (I) model: characterized by the constant current over the input voltage to the load even if the voltage of the node is changed, but by changing the voltage of the node, the power load changes. In this model the load power consumption is proportional to the size of the load voltage varies.

Constant Impedance model (Z): In this model, if the node voltage is changed, load impedance remains constant. Through specification is used to gain power consumption. The power consumption of these groups varies with the square of the load voltage.Constant Power Model (P&Q): In this model if the node voltage is changed, the load power remains constant and also the change in level of node voltage, the current changes is in such a way that an input power supply is needed.

Exponential load model: the load power consumption varies exponentially with voltage measurements.

$$P = P_0 * V^{K_1} \qquad (8)$$

$$Q = Q_0 * V^{K_2} \qquad (9)$$

The amount of active power and reactive P0 and Q0 is at the nominal voltage and V is the voltage in the per unit system. K2 and K1 are also not required to be equal and the amount of active power and reactive power can be sensitive to voltage changes at different values .Coefficients K1 and K2 are not available due to lack of information for most loads characteristics and the distribution companies do not have information about the type of loads, so the loads have been classified into four groups and then the values of K2 and K1 are the selective use.

In this case the coefficient K2 and K1 for each group according to Table 2 will be selected.

TABLE 2: K1 AND K2 VALUES FOR THE FOUR CATEGORIES OF CONVENTIONAL DISTRIBUTION LOADS

| | |
|---|---|
| Constant Load Power | $K2 = K1 = 0$ |
| Constant Current (I) | $K2 = K1 = 1$ |
| Constant Impedance Model (Z) | $K2 = K1 = 2$ |
| Exponential Model | $K2 = 3.22, K1 = 1.38$ |

K1 and K2 values for each load based on using data from laboratory or experimental curve is obtained. For the sample, Table 3 determines the coefficient K1 and K2 based on the classification of load consumptions.

TABLE 3: K1 AND K2 VALUES FOR THE CLASSIFICATION OF LOAD CONSUMPTIONS

| classification of load consumptions | K1 | K2 |
|---|---|---|
| Industrial | 0.18 | 6 |
| Residential | 0.92 | 4.04 |
| Commercial | 1.51 | 3.04 |

c) *The Bus Voltage Limit:*

For proper functioning of equipment and distribution system efficiency, the bus voltage should be in allowed range. In a natural system, the maximum and minimum voltage must not exceed 10% ± range. Thus, using the equation (11), the excessive increase or decrease of voltage is prevented. Equation (10) is the voltage of sub-distribution system that is used as the reference voltage and the voltage value depends on no other buses.

$$V_{pu}(1,t) = V_{pu,Slack,Constant} \qquad (10)$$

$$V_{pu}(i)_{min} \leq V_{pu}(t,i) \leq V_{pu}(i)_{max} \qquad (11)$$

Since load of each buss is sensitive, acceptable efficient voltage range of the amount permitted to be located at each bus voltage of each bus in the system unit can be defined differently.

d) *Capacitance Constraint*

One of conventional and low-cost ways of providing reactive power is using capacitor. Thus in this section this model is discussed. The capacitor which is connected to the network is of type the parallel, and capacitor banks can be formed with a base value with a few capacitors. The amount of reactive power injected by the capacitor bank into the network in this model the equation (12) can be used.

$$Q_{pu,cap} = -V_{pu}(i,t)^2 (2f\pi) \left(NCap(i,t) C_{pu,Base}(i)\right) \qquad (12)$$

The number of capacitors in network at any time shall not be more than the total capacitance of capacitor bank, and also the correct numerical value must be an integer and nonnegative. Equation (13) and (14) are the expression of these constraints.

$$0 \leq NCap(i,t) \leq NCap_{max}(i) \qquad (13)$$

$$NCap(i,t) : Integer \qquad (14)$$

### III. The software Solution

Nature of functions and equations in this model are nonlinear models with integer variables, and to obtain the optimal solution two methods can be used. The first approach is to make the functions linear. Then the use of conventional methods of mathematical operational research and the second method is selecting the appropriate method for solving such a problem like this.

Since to make the functions and equations linear which are used in this non-linear model, is complicated, and also is associated with reduced accuracy of the method for solving the proposed model, the second approach is used to solve the problem, so GAMS software has been selected for solving this problem.

## IV. Sampling

The results of the model in this section go to the part of electricity distribution network feeder of Khodabandelu district in Tehran. Single-line diagram in Figure 1 is depicted.

In this network a 20 kV feeder with 13 buses from substation 20/63 KV is studied. Impedance profile lines are shown in Table 3. This information network has been extracted from reference (Khaje Kazeruni 2001).

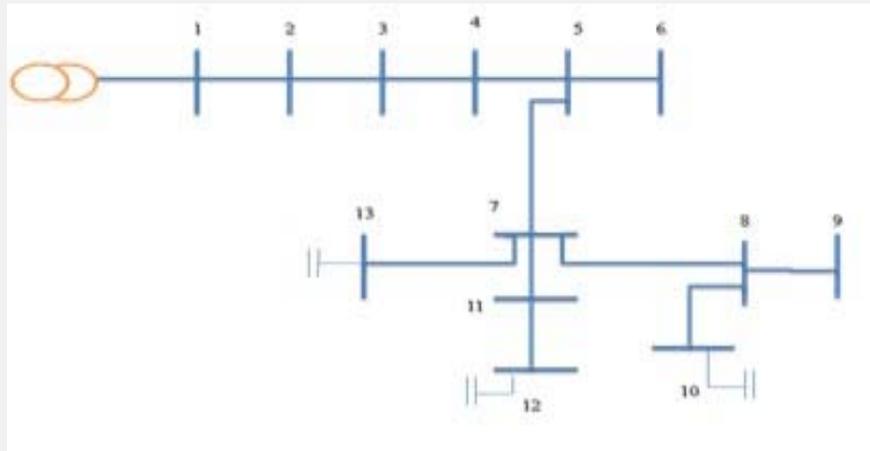

Figure 1. Single line diagram of the distribution network

Also, 3 capacitors on the buses 10, 12 and 13 are considered, and the profile of them in the distribution network in Table 6 is shown. Due to the unavailability of electrical energy consumption per hour at each bus of a feeder, and also according to the fact that in which classification (industrial, residential and commercial) the load is defined, the amount of energy consumption in the distribution network has been studied in two cases. Line characteristic in Table 4 and Classification of the loads and the peak consumption in Table 5, and Table 6 are considered.

TABLE 4: THE LINE CHARACTISTICS (IMPEDANCES)

| From | To | R ohm | X ohm |
|---|---|---|---|
| 1 | 2 | 0.176 | 0.138 |
| 2 | 3 | 0.176 | 0.138 |
| 3 | 4 | 0.045 | 0.035 |
| 4 | 5 | 0.089 | 0.069 |
| 5 | 6 | 0.045 | 0.035 |
| 5 | 7 | 0.116 | 0.091 |
| 7 | 8 | 0.073 | 0.073 |
| 8 | 9 | 0.074 | 0.058 |
| 8 | 10 | 0.093 | 0.093 |
| 7 | 11 | 0.063 | 0.05 |
| 11 | 12 | 0.068 | 0.053 |
| 7 | 13 | 0.062 | 0.053 |

In the first case, all the consumers in the network are considered with a constant power and in the second case; each consumer is modeled with its corresponding load classification.

TABLE 5: THE LOAD CONSUMPTION CLASSIFICATION AND PEAK LOAD

| | Peak Load | | |
|---|---|---|---|
| | P (pu) | Q (pu) | Type of Customers |
| ni2 | 0.089 | 0.0468 | Industrial |
| ni3 | 0.0628 | 0.047 | Industrial |
| ni4 | 0.111 | 0.0767 | Industrial |
| ni5 | 0.064 | 0.0378 | Industrial |
| ni6 | 0.047 | 0.0344 | Residential |
| ni7 | 0.134 | 0.1078 | Industrial |
| ni8 | 0.092 | 0.0292 | Commercial |
| ni9 | 0.077 | 0.0498 | Commercial |
| ni10 | 0.066 | 0.048 | Commercial |
| ni11 | 0.069 | 0.0186 | Residential |
| ni12 | 0.129 | 0.0554 | Commercial |
| ni13 | 0.112 | 0.048 | Residential |

TABLE 6: THE PERCENTAGE OF PEAK CONSUMPTION OF ENERGY IN REGARDING WITH EACH CLASSIFICATION

| T | Commercial | Industrial | Residential |
|---|---|---|---|
| T1 | 10 | 40 | 70 |
| T2 | 2 | 40 | 50 |
| T3 | 2 | 40 | 30 |
| T4 | 2 | 40 | 30 |
| T5 | 2 | 40 | 30 |
| T6 | 10 | 40 | 50 |
| T7 | 10 | 40 | 50 |
| T8 | 20 | 100 | 60 |
| T9 | 60 | 100 | 60 |
| T10 | 90 | 100 | 60 |
| T11 | 100 | 100 | 60 |
| T12 | 100 | 100 | 70 |
| T13 | 100 | 100 | 70 |
| T14 | 100 | 60 | 80 |
| T15 | 100 | 60 | 80 |
| T16 | 90 | 100 | 80 |
| T17 | 90 | 100 | 70 |
| T18 | 90 | 100 | 70 |
| T19 | 100 | 100 | 70 |
| T20 | 100 | 100 | 90 |
| T21 | 100 | 100 | 100 |
| T22 | 90 | 60 | 100 |
| T23 | 80 | 60 | 100 |
| T24 | 30 | 40 | 90 |
| T25 | 10 | 40 | 70 |

TABLE 7: THE SIZE AND NUMBER OF CAPACITOR IN DISTRIBUTION NETWORK

|      | Size of Capacitor (µF) | No |
|------|------------------------|----|
| ni10 | 2                      | 3  |
| ni12 | 2                      | 3  |
| ni13 | 2                      | 3  |

## V. Conclusion

More accurate model of the load consumption continues to increase the accuracy of demand the model but also, run time to solve the model by the software will be prolonged. This model can be used to compare the effect of the load demand of the market for wholesale distribution companies or sub-distribution's bus.